\documentstyle[aps,prb,epsf,twocolumn]{revtex}

\begin{document}

\narrowtext


\title{Impurity corrections to the thermodynamics in spin chains
using a transfer-matrix DMRG method}

\author{Stefan Rommer\cite{address} and Sebastian Eggert}

\address{Institute of Theoretical Physics\\ Chalmers University of
Technology\\ S-41296 G\"{o}teborg, Sweden}


\maketitle

\begin{abstract}
We use the density matrix renormalization group (DMRG) for transfer
matrices to numerically calculate impurity corrections to
thermodynamic properties.  The method is applied to two impurity models
in the spin-1/2 chain, namely a weak link in the chain and an external 
impurity spin. The numerical analysis confirms the field theory calculations 
and gives new results for the crossover behavior.
\end{abstract}

\footnotetext{\copyright The American Physical Society}

\section{Introduction}

The study of quantum impurities remains a large part of condensed matter
physics.  The Kondo effect is maybe one of the most famous examples for
impurity effects, but more recently much effort has been devoted to
impurities in low-dimensional magnetic systems in connection with high
temperature superconductivity\cite{elbio}.  For the particular case of
impurities in quasi one-dimensional systems, much progress has been made
with field theory descriptions e.g.~for the Kondo model\cite{affleck},
quantum wires\cite{kanefisher}, and spin chains\cite{eggert1}.  In those
cases, the impurity behaves effectively as a boundary condition at low
temperatures and the behavior can be described in terms of a renormalization
crossover between fixed points as a function of temperature\cite{affleck}.

Numerically this renormalization picture has been 
confirmed\cite{eggert1,xxz,zhang3}, but 
so far it was only possible to examine a limited number of
energy eigenvalues in the spectrum.  While some efforts have
been made to extract thermodynamic properties from the 
energy spectrum directly\cite{zhang}, such an approach is tedious and 
remains limited by finite system sizes.  
Monte Carlo simulations appear to be well suited for determining
thermodynamic properties, but for the particular case of impurity 
properties it turns out to be difficult to accurately determine 
a correction which is of order $1/N$, where $N$ is the system size.
We now apply the {\it transfer matrix} DMRG to impurity systems. This
overcomes those problems by explicitly taking the thermodynamic limit $N\to
\infty$,  
while still being able to probe impurity corrections and local
properties at finite temperatures even for frustrated systems
(which are not suitable for Monte Carlo simulations due to the 
minus sign problem).  

There are two separate impurity effects that we wish to address.
The first is the impurity correction $F_{\rm imp}$
to the total free energy of a one dimensional system
\begin{equation}
F_{\rm imp} \ = \ \lim_{N\to \infty}(F_{\rm total} - N F_{\rm pure}),
\end{equation}
where $F_{\rm pure}$ is the free energy per site for an infinite
system without impurities.  In other words, the impurity contribution
is that part of the total free energy that does not scale with the 
system size $N$
\begin{equation}
F_{\rm total} = N F_{\rm pure} + F_{\rm imp} + {\cal O}(1/N). \label{free}
\end{equation}
Therefore, the impurity free energy $F_{\rm imp}$ is directly proportional to
the impurity density of the system, and it
immediately determines the corresponding impurity
specific heat and impurity susceptibility, i.e.~quantities that 
can be measured by experiments as a function of temperature and impurity  
density.  Despite the obvious importance of
this impurity contribution we are not aware of any numerical studies
that considered this quantity for any non-integrable impurity
system. Traditional methods would require  
an extensive finite size scaling analysis to track down the $1/N$ correction
to the total free energy per site, but our approach allows us to 
calculate $F_{\rm imp}$  directly in the thermodynamic limit.
We would like to point out that in other studies the
response of an impurity spin to a {\em local} magnetic field is often termed
``impurity susceptibility'', but we prefer to reserve this expression
for the impurity contribution 
\begin{equation}
\chi_{\rm imp} = - \frac{\partial^2}{\partial B^2}{ F_{\rm imp}}, \label{chiimp}
\end{equation}
where $B$ is a {\it global} magnetic field on the total system.

The second aspect of impurity effects are {\it local} properties
of individual sites near the impurity location, e.g.~correlation 
functions and the response to a local magnetic field.  Local
properties have been the central part of a number of works for
many impurity models\cite{xxz,zhang3,zhang,zhang2,igar,NMR}.  
Our approach is now able to calculate 
these impurity effects directly in the thermodynamic limit and we
get quick and accurate results to extremely low temperatures.
It turns out that the local impurity effects can be 
determined much more accurately and to lower temperatures
than the impurity contribution $F_{\rm imp}$, which remains 
limited by accuracy problems even with this method. 

The Density Matrix Renormalization Group (DMRG)\cite{White92} has had a
tremendous success in describing low energy static properties of many 
one-dimensional (1D) quantum systems such as spin chains and electron 
systems. More recently many useful extensions to the DMRG have been
developed. Nishino showed how to successfully apply the density matrix idea to
two-dimensional (2D) classical systems by determining the largest eigenvalue
of a transfer matrix\cite{Nishino95}. Bursill, Xiang and Gehring have
then shown that the same idea can 
be used to calculate thermodynamic properties of the quantum 
spin-1/2 $XY$-chain\cite{Bursill96}. 
The method has later been
improved by Wang and Xiang\cite{Wang97} as well as
Shibata\cite{Shibata97} and been applied to the 
anisotropic spin-1/2 Heisenberg chain with great success. In this paper we
apply a generalization of the method to impurity systems, which is 
presented in section \ref{method}.
In Section III we study two different 
impurity models in the spin-1/2 chain and are able to 
confirm predictions from field theory calculations.
Section IV concludes this work with a discussion of the results
and a critical analysis of accuracy and applicability to other systems.

\section{The Method} \label{method}
The method of the transfer matrix DMRG can in principle be applied to any 
one-dimensional system for which a transfer matrix can be defined.
As a concrete example we will consider the antiferromagnetic 
spin-1/2 chain, since this model is well understood in terms of field 
theory treatments and has direct experimental relevance.
The Heisenberg Hamiltonian can be written as
\begin{equation}
H = \sum_{i=1}^{N} h_i, \; \; \; \; h_i = J_i \; {\bf S}_i \cdot {\bf
S}_{i+1} + B_i \; S_i^z , \label{eq:ham}
\end{equation}
where $J_i$ is the exchange coupling between sites $i$ and $i+1$, and $B_i$
is an external  magnetic field in the $z$-direction at site $i$. 
Periodic boundary conditions, ${\bf S}_{N+1} \equiv {\bf S}_1$, are
assumed. The partition function is defined by 
\begin{equation}
Z = \mbox{tr} \; e^{- \beta H} = \mbox{tr} \; e^{- \beta (H_o + H_e)} ,
\end{equation}
where $\beta = \frac{1}{k_B T}$ and where we in the last step have
partitioned $H$ into odd and even site terms,
\begin{equation}
H_o = \sum_{i=1}^{N/2} h_{2i-1}, \; \; \; \;  H_e = \sum_{i=1}^{N/2} h_{2i} .
\end{equation}

\subsection{The transfer matrix method}
The quantum transfer matrix for this system is defined as usual via
the Trotter-Suzuki decomposition\cite{Trotter59} 
\begin{equation}
Z_M = \mbox{tr} \left( e^{- \frac{\beta H_o}{M} }
e^{-\frac{\beta H_e}{M}} \right)^M , \label{eq:trot} 
\end{equation}
where $M$ is the Trotter number. This expression approximates the partition
function up to an error of order $(\beta/M)^2$ and becomes exact in the
limit $M \rightarrow \infty$.
By inserting a complete set of states between each of the exponentials in
Eq.~(\ref{eq:trot}) and rearranging the resulting matrix elements, the
partition function can be written as a trace over a product of transfer
matrices,\cite{Bursill96}
\begin{equation}
Z_M = \mbox{tr} \; \prod_{i=1}^{N/2} {T}_M({2i-1}) , \label{eq:zm}
\end{equation}
where ${T}_M({2i-1})$ is the $2^{2M} \times 2^{2M}$ dimensional quantum
transfer matrix from lattice site $2i-1$ to site $2i+1$.
Note that ${T}_M$ is in general not symmetric. However, if the two-site
Hamiltonian $h_{2i-1}$ and $h_{2i}$ of Eq.~(\ref{eq:ham}) is invariant under
the exchange $2i-1 \leftrightarrow 2i$, and $2i \leftrightarrow 2i+1$
respectively, as is the case unless we have applied
a non-uniform magnetic field, ${T}_{M}$ is a product of two
symmetric transfer matrices, one from site $2i-1$ to site $2i$ and the other
from site $2i$ to $2i+1$.  
For a uniform system, the transfer matrix is independent of lattice site,
${T}_M({2i-1}) \equiv {T}_M$,  and the partition function is then
given by
\begin{equation}
Z_M = \mbox{tr} \; {T}_M^{N/2} .
\end{equation}
In the thermodynamic limit of the uniform system, $Z_M$ is given by
\begin{equation}
\lim_{N \rightarrow \infty} Z_M = \lambda^{N/2} , \label{eq:zlim}
\end{equation}
where $\lambda$ is the largest eigenvalue of ${T}_M$. 

The largest eigenvalue $\lambda$ can be found exactly only for small
Trotter numbers $M$. As $M$ increases, the dimension of ${T}_M$ grows
exponentially, and we have to use some approximation technique to 
find $\lambda$. Analogous to the case where the DMRG can be used to
find a certain eigenstate of a Hamiltonian as the number of lattice sites
increase, we can use the DMRG to find the largest eigenvalue of ${T}_M$
as the Trotter number $M$ increase. The strategy is thus to start with a
system block, ${T}^s_{M/2}$, and an environment block, ${T}_{M/2}^e$, with a 
small $M$. The superblock transfer matrix, ${T}_{M}$, with Trotter
number $M$, is constructed by ``gluing'' together the system block with the
environment block. Periodic boundary conditions in the Trotter direction must 
be used. The reduced density matrix is constructed from the target state,
i.e.\ the eigenstate of ${T}_{M}$ with largest eigenvalue. Since the
transfer matrix is non-symmetric, the left and right eigenvectors will not
be complex conjugates of each other. 

A reduced density matrix for the system as part of the superblock can be
constructed by taking a partial trace of $T_M^{N/2}$ over the environment
degrees of freedom\cite{Wang97},
\begin{equation}
\rho = \frac{1}{Z_M} \mbox{tr}_{\rm env} T_M^{N/2} . \label{eq:reddens}
\end{equation}
In the thermodynamic limit only the state with the largest eigenvalue will
contribute, 
\begin{equation}
\rho \stackrel{N \to \infty}{\longrightarrow} \mbox{tr}_{\rm env} | \psi^R
\rangle \langle \psi^L | , \label{eq:limdens} 
\end{equation}
where $| \psi^R \rangle $ and $ \langle \psi^L |$ are the right 
and left eigenvectors of the superblock transfer matrix, $T_M$,
corresponding to the largest eigenvalue, $\lambda$. The matrix elements are
given by 
$\rho_{i^\prime,i} = \sum_j \psi^R_{i^\prime,j} \psi^L_{i,j}$ , where $i$
and $j$ label the degrees of freedom of the system and the 
environment respectively and the target states are given by $ \langle \psi^L
|  = \sum_{i,j} \psi^L_{i,j} \langle i | \langle j | $ and $ | \psi^R
\rangle = \sum_{i,j} \psi^R_{i,j} | i \rangle | j \rangle $. 

The left and right eigenvectors of the density matrix with the largest
eigenvalues are then used to define the projection operators onto the
truncated basis.  
After the first iterations we will keep $m$ states for the
system and the environment and the superblock will be $4m^2$ dimensional. 
More details on the transfer matrix DMRG algorithm for quantum systems have
been presented in Refs.~\onlinecite{Bursill96,Wang97,Shibata97}.

Since $\rho$ is non-symmetric it is not obvious that the eigenvalues are
real, but because $\rho$ represents a density matrix we expect it to be
positive definite. However, because of numerical inaccuracies 
 complex eigenvalues tend to appear in the course
of iterations. This is usually connected to level crossings in
the eigenvalue spectra of the density matrix as $M$ is increased. 
In addition, we also observe that just before the complex eigenvalues
appear, 
multiplet symmetries in the eigenvalue spectrum of $\rho$ may split
up (often due to ``level repulsion'' from lower lying states which were
previously neglected).  To overcome this problem
it is important to keep all original symmetries (i.e.~eigenvalue
multiplets of $\rho$) even as $M$ increases.  For that purpose it is often
necessary to increase the number of states $m$ 
just before a multiplet tends to split up, which avoids the numerical error
that leads to the symmetry breaking. In case of a level crossing of two
multiplets that do not split up, complex eigenvalues may still appear and
it is then possible to numerically transform the complex
eigenstate pair into a real pair spanning the same space. Since the
transformed real pair still is orthonormal to every other eigenvector, this
transformation does not cause any troubles for later iterations. With this
method the eigenvalues stay complex until the two levels have crossed and
moved enough apart, at which point the eigenvalues become real again. The
renormalization procedure is therefore not as straightforward as in ordinary
DMRG runs, since it is essential to track {\it all} eigenvalues of $\rho$
and to dynamically adjust the number of states $m$, thereby
sometimes repeating previous iteration steps.

\subsection{Impurities and local properties}
The renormalization scheme also allows 
the calculation of thermal expectation values of local operators.
The magnetization of the spin at site $i$ is determined by
\begin{equation}
\langle S_i^z \rangle = \frac{1}{Z} \; \mbox{tr} \; S_i^z e^{-\beta H} .
\end{equation}
The operator $S_i^z$ thus only has to be incorporated in the corresponding
transfer matrix at site $i$. For the {\it pure} system, 
we then arrive at the formula\cite{Wang97}
\begin{equation}
\langle S_i^z \rangle = \frac{\langle \psi^L | {T}_M^{sz}(i) | \psi^R 
\rangle}{\lambda} , 
\end{equation}
where ${T}_M^{sz}(i)$ is defined similar to ${T}_M(i)$ but with the
additional operator $S^z_i$ included in addition to the Boltzmann weights. 
To measure the local bond energy, $h(i) = {\bf S}_i \cdot {\bf S}_{i+1}$, a
similar construction can be used.

\begin{figure}
\epsfxsize=3.35in
\epsfbox{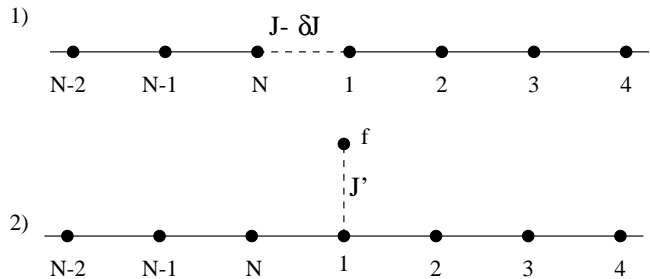}
\caption{The impurity configurations considered in 
\protect{Eqs.~(\ref{imp1}) and (\ref{imp2})}. Note, that
the left and right ends are joined with periodic boundary conditions.}
\label{fig:impsyst}
\end{figure}

Let us now assume that the system has a single impurity. The systems we will
study is the periodic spin-1/2 chain with one weakened link or
an external spin-1/2 coupled to one spin in the chain
\begin{eqnarray}
H_1 & = & H_0 -  \delta \! J \; {\bf S}_{N} \cdot {\bf S}_1  \label{imp1}\\ 
H_2 & = & H_0 +  J^\prime \; {\bf S}_1 \cdot {\bf S}_f, \label{imp2}
\end{eqnarray}
where 
\begin{equation}
H_0 = \sum_{i=1}^{N-1} J {\bf S}_{i} \cdot {\bf S}_{i+1} + 
J {\bf S}_{N} \cdot {\bf S}_1
\end{equation} is the {\it periodic} chain and
 ${\bf S}_f$ is the spin 
operator of an external spin-1/2. The models are depicted in
Fig.~\ref{fig:impsyst}. 

For systems with such a local impurity, which is contained within two
neighboring links, only one of the $T_M(i)$ in Eq.~(\ref{eq:zm}) will differ
from a common $T_M$
\begin{equation}
Z_M = \mbox{tr} \; \left( T_M^{N/2-1} T_{\rm imp} \right) , \label{eq:zimp} 
\end{equation}
where $T_{\rm imp}$ is the transfer matrix of the two links containing the
impurity and $T_M$ is the transfer matrix describing the bulk.
In the thermodynamic limit the partition function will still be dominated by
the largest eigenvalue of the ``pure'' transfer matrix. From
Eq.~(\ref{eq:zimp}) we have
\begin{equation}
\lim_{N \rightarrow \infty} Z_M = \lambda^{N/2-1} \langle \psi^L | {T}_{\rm
imp} | \psi^R \rangle , \label{eq:implim}
\end{equation}
where $\lambda$, $\langle \psi^L | $ and $| \psi^R \rangle $ all correspond
to the pure system. 
The generalization of Eqs.~(\ref{eq:zimp}) and (\ref{eq:implim}) to
impurity configurations ranging over more than two links is
straightforward. In this case more 
than one impurity transfer matrix has to be introduced and this could be
used to study e.g.\ multiple impurities and impurity-impurity interactions.  

Let us define $\lambda_{\rm imp}\equiv \langle \psi^L | {T}_{\rm imp}
| \psi^R \rangle$. The total free energy of the system is then given by
\begin{eqnarray}
F & = & -\frac{1}{\beta} \ln Z \nonumber \\
& = & -\frac{1}{\beta} \ln \left( \lambda^{N/2-1} \lambda_{\rm imp} \right) 
\nonumber \\ 
& = & -\frac{N}{2 \beta} \ln \lambda - \frac{1}{\beta} \ln
\frac{\lambda_{\rm imp}}{\lambda} .
\end{eqnarray}
By comparing with Eq.~(\ref{free}) we can retrieve the pure and impure parts
\begin{equation}
F_{\rm pure} = -\frac{1}{2 \beta} \ln \lambda , \; \; \; F_{\rm imp} =
-\frac{1}{\beta} \ln \frac{\lambda_{\rm imp}}{\lambda} . \label{impF}
\end{equation}
The impurity susceptibility can be found from the change of $F_{\rm imp}$ in a
small magnetic field from Eq.~(\ref{chiimp})
\begin{equation}
\chi_{\rm imp} = -\frac{\partial^2 }{\partial B^2}{ F_{\rm imp}}
.\label{chiimp2} 
\end{equation}
Local properties such as the magnetization of the impurity spin can be
determined by
\begin{equation}
\langle S_{\rm imp}^z \rangle = \frac{ \langle \psi^L | {T}_{\rm imp}^{sz} |
\psi^R \rangle }{ \lambda_{\rm imp} } .
\end{equation}
The magnetization of spins close to the impurity is readily obtained
by 
\begin{equation}
\langle S^z \rangle = \frac{ \langle \psi^L | {T}_{M}^{sz} ({T}_{M})^x
{T}_{\rm imp} | \psi^R \rangle }{ \lambda^{x+1} \; 
\lambda_{\rm imp} } , \label{eq:szclose} 
\end{equation}
where $2x$ is the number of sites between the impurity and the
spin of interest. Note that since a transfer matrix involves a total of three
lattice sites, ${T}^{sz}_M$ can be constructed to measure the spin at any
of these sites (or the mean value). The actual site of the measurement in 
Eq.~(\ref{eq:szclose}) is thus determined both by $x$ and how ${T}^{sz}_M$
is set up.  The expectation value in Eq.~(\ref{eq:szclose}) is most easily 
calculated by first computing the 
vectors $\langle \psi^L | {T}_{M}^{sz}$ and ${T}_{\rm imp} | \psi^R
\rangle$, then acting with ${T}_{M}$ on one of these states, and finally
calculating the inner product of the resulting states.
Eq.~(\ref{eq:szclose}) can be generalized to measure any
equal time correlation function with or without an impurity, e.g.~by
replacing ${T}_{\rm imp}$ by ${T}_{M}^{sz}$.

The reduced density matrix for the impurity  system  can be constructed by
taking the thermodynamic limit of the impurity version of
Eq.~(\ref{eq:reddens}), 
\begin{equation}
\rho = \frac{1}{Z_M} \mbox{tr}_{\rm env} \left( T_M^{N/2-1} T_{\rm imp}
\right) , \label{eq:impdens} 
\end{equation}
with $Z_M$ as in Eq.~(\ref{eq:zimp}). In our calculations we
have used the same density matrix as for the pure case, i.e.\
Eq.~(\ref{eq:limdens}), and we have found it to give good results. This form
can most easily be motivated by writing Eq.~(\ref{eq:impdens}) on the form
$\rho = \mbox{tr}_{\rm env} \left( T_M^{N/4-1} T_{\rm imp} T_M^{N/4}
\right) / Z_M $. 
From a computational point of view, this method is also very convenient; by
storing all target states and projection operators from the 
DMRG run for the {\it pure} system, all local impurities can be studied by
simply using the same projection operators and target states. This makes
subsequent DMRG runs for different impurity parameters very fast.

There are also other choices of density matrices that can be
made. The thermodynamic limit of Eq.~(\ref{eq:impdens}) can also be
interpreted as  $\rho \rightarrow | \psi^R
\rangle \langle \psi^L | T_{\rm imp}/\lambda_{\rm imp}$ in which case the 
impurity transfer matrix would be taken into account in the density
matrix. This would in some sense be analogous to including an operator
different from the 
Hamiltonian in the density matrix of the ordinary ``zero-temperature''
DMRG,  which is usually not necessary to measure e.g.\ correlation
functions in the ordinary DMRG. 
This approach would destroy the computational advantage of using 
the pure density matrix, because the pure projection operators and target
states could not be used but instead complete DMRG runs would
have to be done for each impurity configuration and coupling.

\section{Results}
\subsection{Field theory predictions}
To make a meaningful analysis of the numerical results, we first 
need to understand the spin-1/2 chain in the framework of the quantum 
field theory treatment.  This turns out to give a good description
of the impurity behavior in terms of a renormalization flow between 
fixed points, and we will be able to set up concrete expectations
for the impurity susceptibility in Eq.~(\ref{chiimp}) as well as
local properties. 

The effective low energy spectrum of the spin-1/2 chain is 
well described by a free boson Hamiltonian density
\begin{equation}
{\cal H} = \frac{v}{2} \left[ (\Pi_\phi)^2 \,
+ \, ({\partial_x \phi})^2 \right],  \end{equation}
plus a marginal irrelevant operator $\cos \sqrt{8 \pi} \phi$ 
and other higher order operators which we have neglected.
Here $\Pi_{\phi}$ is the momentum variable conjugate to $\phi$.
In the long wave-length limit the spin operators can be expressed in
terms of the bosonic fields using the notation of Ref.~\onlinecite{eggert1}
\begin{eqnarray} S^z_j &\approx & {1\over\sqrt{ 2\pi}}{\partial \phi \over
\partial x} +(-1)^j\hbox{const.}\ \cos {\sqrt{ 2\pi}\phi } \nonumber \\
S^-_j &\approx & e^{i\sqrt{2\pi} \tilde \phi}[\hbox{const.}\, 
\cos {\sqrt{ 2\pi}\phi} + (-1)^j\hbox{const.}],
\label{cont-op}\end{eqnarray}
At this point we can introduce the impurities in Eq.~(\ref{imp1})
and (\ref{imp2}) in a straightforward way as perturbations.
The field theoretical expressions for these perturbations can then
be analyzed in terms of their leading scaling dimensions. 
Local perturbations with a scaling dimension of $d>1$ are considered
irrelevant, while perturbations with a small scaling dimension
$d < 1$ are relevant and drive the system to a different fixed point.
Hence, we can predict a
systematic renormalization flow towards or away the corresponding
fixed point, respectively.  

Such an analysis has been made in Ref.~\onlinecite{eggert1} and the 
renormalization flows have been confirmed by determining the finite size 
corrections to the low energy spectrum\cite{eggert1}.  
In particular, a small weakening of one link in the chain
\begin{equation}
H_1  =  H_0 -  \delta \! J \; {\bf S}_{N} \cdot {\bf S}_1
\end{equation}
has been found to be a relevant perturbation described by the 
operator $\sin \sqrt{2 \pi} \phi$ with scaling dimension $d=1/2$, so 
that the periodic chain $(\delta \! J=0)$ is an unstable fixed point.  The
open chain  $(\delta \! J=J)$ on the other hand is a stable fixed point where the
perturbation is described by the leading irrelevant operator
 $\partial_x \phi(N)\partial_x \phi(0)$ with scaling dimension of $d=2$.
Hence we expect a renormalization flow between the two fixed points
as the temperature is lowered, and the temperature dependence of
the impurity susceptibility as well as local properties will be described 
by a crossover function.  Below a certain crossover temperature $T_K$
this crossover function describes the behavior of the stable
fixed point (the open chain) while above $T_K$ the system may exhibit
a completely different behavior.  The crossover temperature $T_K$ itself
is determined by the initial coupling strength $\delta \! J$ 
\begin{eqnarray}
\lim_{\delta \! J \to 0} T_K &  \to &  0 \nonumber \\
\lim_{\delta \! J \to J} T_K  & \to & \infty. \label{T_K} \end{eqnarray}
In other words, close to the unstable fixed point the crossover temperature 
is very small, indicating that we have to go to extremely low temperatures
before we can expect to observe the behavior of the stable fixed point.

A similar scenario holds for the impurity model with 
one external spin ${\bf S}_f$,
\begin{equation}
H_2  =  H_0 +  J' \; {\bf S}_{1} \cdot {\bf S}_f.
\end{equation}
In this case, the periodic chain $(J'=0)$ is also the unstable fixed point,
while the open chain with a decoupled singlet ($J' \to \infty$)
is the stable fixed point\cite{eggert1,xxz}.

As mentioned above, the renormalization flow has been confirmed for
the energy corrections of individual eigenstates, but we now
seek to extend this analysis to thermodynamic properties, which 
will allow us to determine the crossover behavior and $T_K$ directly.

\subsection{One weak link}
\begin{figure}
\epsfxsize=3.35in
\epsfbox{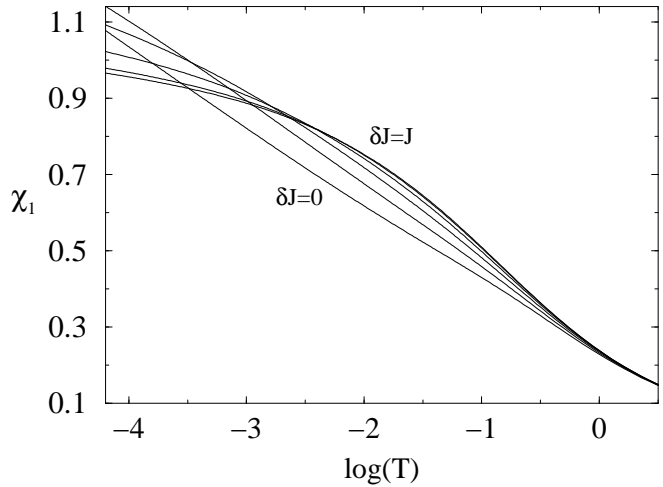}
\caption{The local response $\chi_1$ of the spin ${\bf S}_1$ next to 
a weak link for different coupling strengths
$\delta \! J/J= 0, 0.2,  0.4, 0.6,  0.8,  1$.
The crossover to the open chain behavior occurs at different temperatures
$T_K$ depending on the coupling strength.}
\label{chi1}
\end{figure}

Our first task is to establish the renormalization behavior
from a periodic chain fixed point to the open chain fixed point
as a function of temperature. To examine the effective boundary 
condition on the spin-operators, it is instructive to look at the 
correlation functions at spin-sites close to the impurity.
For periodic boundary conditions, the leading operator for the  
spin $S^z_1$-operator is given  by $\cos {\sqrt{ 2\pi}\phi }$ with
scaling dimensions of $d = 1/2$ according to Eq.~(\ref{cont-op}).
On the other hand, open boundary conditions restrict the allowed operators
and the leading operator for $S^z_1$ is found to be $\partial_x \phi(0)$ 
with scaling dimension of $d=1$.  Hence, the autocorrelation 
function at the impurity behaves differently, depending on the 
effective boundary condition
\begin{equation}
\langle S^z_1(\tau) S^z_1(0)\rangle \propto \left\{ \begin{array}{lll}
1/\tau &\phantom{nnn} & {\rm periodic\ b.c.} \\
1/\tau^2 &\phantom{nnn} & {\rm open\ b.c.} 
\end{array} \right.
\end{equation}
Therefore, a useful quantity to consider is the response $\chi_1$
to a {\it local} magnetic field given by the Kubo formula
\begin{eqnarray}
\chi_1 (T) & = & \int^{1/T} \langle S^z_1(\tau) S^z_1(0)\rangle \, d \tau 
\nonumber\\ & 
\stackrel{T\to 0}{\longrightarrow} &\left\{ \begin{array}{lll}
 -\log T&\phantom{nnn} & {\rm periodic\ b.c.} \\
{\rm const.} + {\cal O}(T) &\phantom{nnn} & {\rm open\ b.c.} 
\end{array} \right. \label{eq:chi1}
\end{eqnarray}
In Fig.~\ref{chi1} we have presented the results of $\chi_1$ for
different impurity strengths $\delta \! J$ on a logarithmic temperature scale.
For the periodic chain ($\delta \! J = 0$) we clearly observe the logarithmic 
scaling, but for any finite $\delta \! J$ a turnover to a constant behavior
is observed as $T\to 0$, i.e.~the behavior of the open chain.  
The turnover temperature $T_K$ occurs at larger and larger 
values as we approach the stable fixed point [see Eq.~(\ref{T_K})]. An 
interesting aspect is that the curves actually cross:  
At very high temperatures the high temperature
expansion always dictates a larger response for a weakened link, while
at very low temperatures this relation is reversed by quantum mechanical 
effects and the renormalization flow.

\begin{figure}
\epsfxsize=3.35in
\epsfbox{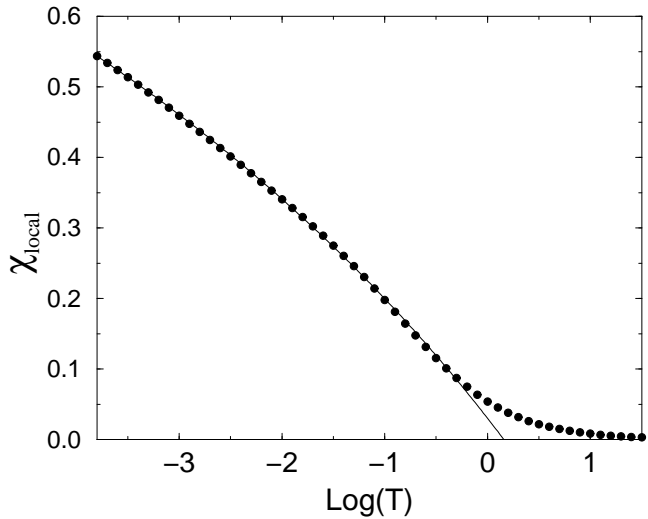}
\caption{The impurity susceptibility for an open chain,
which is approximated by the excess susceptibility at an open end 
$\chi_{\rm local}$ (dots). At moderately low temperatures our data-fit
(solid line) is consistent with the behavior in
\protect{Eq.~(\ref{chiopen})}. The fit is $\chi_{\rm local} =
1.895 + 18.6 / \log (T/T_0)$, with $\log(T_0) = 10$.}
\label{fig:chiopen}
\end{figure}

We now turn to the true impurity susceptibility of Eq.~(\ref{chiimp})
which is the experimentally more relevant quantity.  Maybe the
simplest non-trivial case to consider is the open chain $\delta \! J= J$.  
At low temperatures the
impurity susceptibility can be calculated from the leading irrelevant 
local operator which is allowed in the Hamiltonian.  This operator turns
out to be $(\partial_x \phi(0))^2$ which gives a constant 
impurity susceptibility with a logarithmic correction similar
to the pure susceptibility\cite{eggert3} as follows from a dimensional analysis.
(In fact this operator can 
be absorbed in the free Hamiltonian by a defining  a velocity $v$ that
depends on the system size\cite{affl4}.  
An explicit calculation of integrals
over the correlation functions also comes to the same conclusion.)
\begin{equation}
\chi_{\rm imp}^{\rm open} (T) \ 
\stackrel{T\to 0}{\longrightarrow} \  {\rm const.} + {\cal O}\left(
1/\log(T/T_0)\right)
\label{chiopen}
\end{equation}
While it is possible to calculate the impurity susceptibility
numerically according to Eq.~(\ref{impF}), 
the use of a second derivative in Eq.~(\ref{chiimp2})
causes large problems with the accuracy at lower temperatures
since it involves taking the differences of large numbers.
Luckily, the excess local susceptibility $\chi_{\rm local}$
of the first site under a {\it global}
magnetic field turns out to give a good estimate of the true
impurity susceptibility\cite{NMR}
\begin{equation}
\chi_{\rm local} = \frac{d \langle S^z_1 \rangle}{dB}-\chi_{\rm pure} 
\propto \chi_{\rm imp} + {\rm const.} \label{chilocal}
\end{equation}
where $B$ is a {\it global} magnetic field and the constant is due to 
the alternating part\cite{NMR}.  The results for this quantity are shown 
in Fig.~\ref{fig:chiopen} which are consistent with Eq.~(\ref{chiopen}).  
The second derivative in Eq.~(\ref{chiimp2}) has 
a similar behavior in the intermediate temperature range, but 
is not accurate enough to extrapolate to the $T\to 0$ limit as
explained above.

\begin{figure}
\epsfxsize=3.35in
\epsfbox{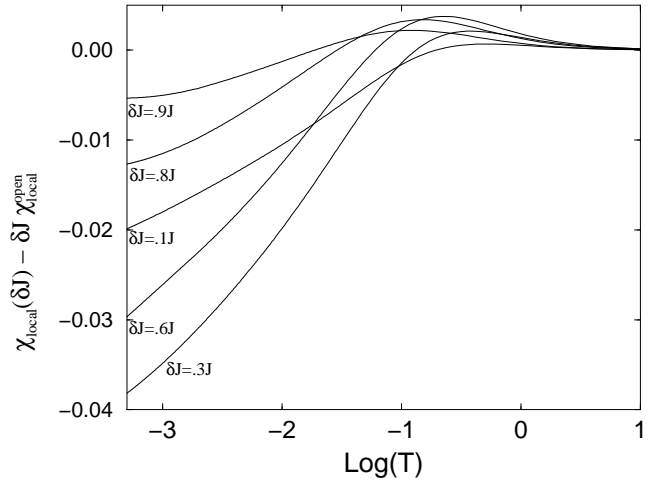}
\caption{The part of the impurity susceptibility which is attributed
to the effects of the weakened link.  For a very weak link  
$\delta \! J \sim J$
the magnitude increases linearly with $J-\delta \! J$
and saturates quickly, while
the magnitude decreases again as we approach the 
periodic chain $\delta \! J \ll J$, but does not saturate as $T\to 0$.}
\label{fig:chiimp}
\end{figure}

Now we are in the position to consider the impurity susceptibility
of one weak link in the chain.
By reducing $\delta \! J$ it is possible to tune the the system all
the way from the open chain fixed point to the periodic chain.
The operator $(\partial_x \phi(0))^2$, which was responsible for the
open chain impurity susceptibility is thereby reduced continuously.
However, it is an entirely different operator corresponding to 
${\bf S}_N\cdot {\bf S}_1$ which is responsible for the 
renormalization.  This operator changes scaling dimension as
we go from periodic boundary conditions 
towards the open chain 
\begin{equation}
{\bf S}_N\cdot {\bf S}_1 \propto \left\{ \begin{array}{lll}
\sin \sqrt{2\pi} \phi \ &d=1/2\  & {\rm periodic\ b.c.} \\
\partial_x \phi(N) \partial_x \phi(0)\  &d=2\  & {\rm open\ b.c.} 
\end{array} \right.
\label{weaklink}
\end{equation}
Since we wish to study the effect of this renormalization, we
choose to subtract the open contribution systematically from 
$\chi_{\rm imp}$, so we obtain exactly the part which will exhibit the
crossover of the renormalization 
\begin{equation}
\chi_{\rm imp} (\delta \! J) - \delta \! J\ \chi_{\rm imp}^{\rm open} 
= f(T/T_K)/T_K. \label{chidiff}
\end{equation}
We see that this difference is zero at either fixed point. 
After subtracting $\delta \! J \chi_{\rm imp}^{\rm open}$
the impurity susceptibility
comes only from the operator in Eq.~(\ref{weaklink}), which allows us to
postulate the  scaling form in Eq.~(\ref{chidiff}). 
Below $T_K$ the difference in Eq.~(\ref{chidiff})
will asymptotically go to a constant for a very weak link $\delta \! J \sim
J$, which comes from the $d=2$ operator in Eq.~(\ref{weaklink}).
Indeed we observe in Fig.~\ref{fig:chiimp} that this 
difference is {\it negative} 
and proportional to $J-\delta \! J \sim 1/T_K$ and largely 
temperature independent as $T \to 0$.  
On the other hand as $\delta \! J$ becomes small enough,
the behavior is much different: The expression in Eq.~(\ref{chidiff})
even decreases  with $\delta \! J$ and the turn-over to constant behavior
happens at much lower temperatures (outside the plot range).

Interestingly, this results in a highly nontrivial behavior as
a function of $T$ and $T_K$ close to the unstable fixed point
\begin{eqnarray}
\lim_{T\to 0} \lim_{T_K \to 0}  f(T/T_K)/T_K  & \propto  & 
T_K/T^2 \ \propto \ \delta \! J^2 \to 0 \nonumber \\
\lim_{T_K\to 0} \lim_{T \to 0}  f(T/T_K)/T_K  & \propto  & 
1/T_K \ \propto \  1/\delta \! J^2 \to -\infty.  \label{limits}
\end{eqnarray}
{\it This means that at $T=0$ a minute perturbation $\delta \! J$ results in
an extremely large negative $\chi_{\rm imp} \propto 1/\delta \! J^2$ } 
(although this behavior occurs in an ``unphysical'' limit).
The reason that the two limits do not commute is of course because
one describes the behavior of the stable fixed point $T \ll T_K$,
while the other one describes the behavior at the unstable fixed point.
While our numerical results cannot show the entire crossover of 
Eq.~(\ref{chidiff}), 
the increase below $T_K$ for small $J-\delta \! J$ is clearly observed
as well as the decrease and change of curvature above $T_K$ for
small $\delta \! J$.
Therefore, our data in Fig.~\ref{fig:chiimp} supports the 
renormalization scenario and the nontrivial behavior of Eq.~(\ref{limits}).

\subsection{One external spin}
\begin{figure}
\epsfxsize=3.35in
\epsfbox{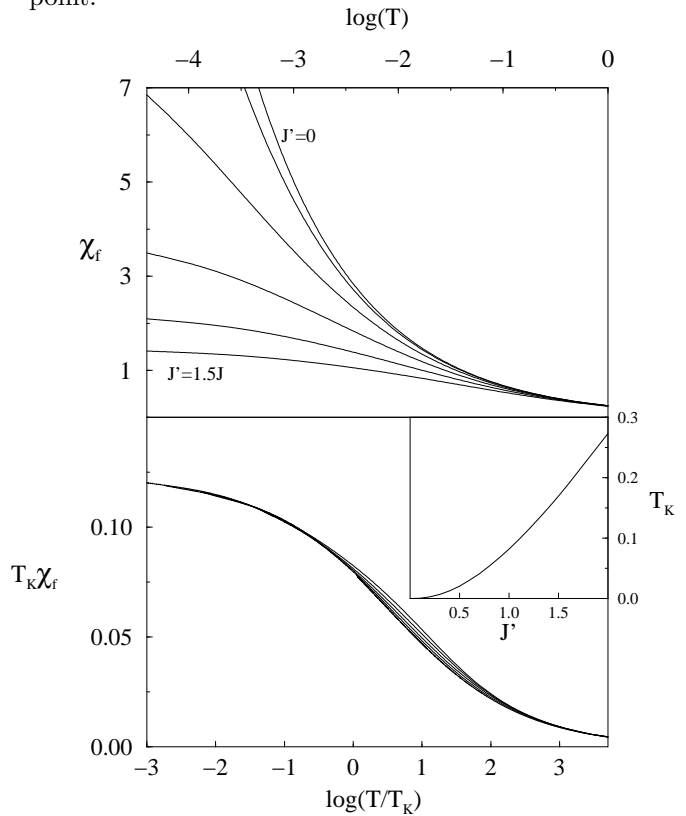}
\caption{ The upper figure shows the local response of the impurity spin
$ \chi_f $ for different coupling strengths $J'/J = 0, 0.3, 0.6, 0.9, 1.2,
1.5$.  The lower figure shows the 
data collapse according to \protect{Eq.~(\ref{chifscaling}).} In the inset
the corresponding crossover temperature $T_K$ is shown, which is
expected to be $T_K \propto {J^\prime}^2$ for $J' \ll J$ and $T_K \propto
J^\prime$ for $J^\prime \gg J$.}
\label{chif}
\end{figure}

The model of an external spin antiferromagnetically coupled to the
chain in Eq.~(\ref{imp2}) is maybe a little more exotic, but is
still of great interest in a number of 
studies\cite{eggert1,xxz,zhang3,zhang,zhang2,igar,liu}.
In Ref.~\onlinecite{eggert1} it was first shown that 
the stable fixed point corresponds to open boundary conditions 
with a decoupled singlet.  This was confirmed numerically\cite{zhang3}, 
but more recently Dr.~Liu postulated a completely different behavior using 
some non-local transformations on Fermion-fields which 
mysteriously were rearranged to form a solvable model\cite{liu}. While we
cannot trust or understand many of his calculations,  the 
predictions are in strong contrast to any previous expectations
and should be tested explicitly.  In particular, he predicted
the response of the impurity spin to a local magnetic field
\begin{equation}
\chi_f (T) = \int^{1/T} \langle S^z_f(\tau) S^z_f(0)\rangle \, d \tau
\end{equation}
to be proportional to $T^{5/2}$ at the Heisenberg point.
We predict, however, that this response is described by the 
autocorrelation function of the leading
operator for ${\bf S}_f$.  By  a symmetry analysis we find that for open
boundary conditions this operator is given by 
$\partial_x \phi(0)$ with scaling dimension $d=1$.  The
local response is therefore a constant as $T\to 0$ with a linear
term
\begin{equation}
\chi_f (T) \stackrel{T\to 0}{\longrightarrow}
{\rm const.} + {\cal O}(T). \label{eq:chif}
\end{equation}
This also agrees with the findings in Ref.~\onlinecite{xxz} which had
similar reservations about Ref.~\onlinecite{liu}.
We now explicitly calculate $\chi_f$ for several coupling strengths $J'$
as shown in Fig.~\ref{chif}. 
Our data fits well to the predicted form and
we can certainly rule out any $T^{5/2}$ behavior.
Moreover, we find a scaling behavior which holds for all coupling strengths
\begin{equation}
\chi_f (T) = g(T/T_K)/T_K .  \label{chifscaling}
\end{equation}
A similar scaling relation was observed before\cite{zhang2}.
Our results are
consistent with previous numerical studies\cite{zhang}, but distinctly larger 
in the low-temperature region. We attribute this to the finite
size method used in Ref.~\onlinecite{zhang}, which 
becomes unreliable when the temperature falls below the finite size gap of 
the system.
At large $J'$ our results can be compared to that of two
coupled spins forming a singlet.
Note, that the response to a local magnetic field on one spin
in a singlet is finite as $T\to 0$ and proportional to 
$1/J'$ (and does {\it not} show activated behavior as a simple 
calculation shows). Therefore, our findings are completely consistent with 
the expectation that the impurity spin is locked into a singlet
at the stable fixed point.

\section{Discussion}

To accurately calculate impurity properties we have to determine not only the
largest eigenvalue of the transfer matrix, ${T}_M$, but
also the corresponding eigenvectors to high accuracy. To estimate the error
of the impurity 
properties is difficult. Errors come both from the finite Trotter number,
$M$, and from the finite number of states, $m$, in the DMRG. The scaling of
pure properties with $M$ and $m$ usually turn out to be simpler than the
scaling of impurity properties, which show a less clear form of the errors.

We have used the value $\beta/M = 0.05$ for all calculations
presented in this article. With this value, the error due to the finite $M$
should be small, which is also confirmed by test-runs. 

We have tested the error due to the finite Trotter number $M$ by doing
separate DMRG runs for different 
values of $\beta/M$. For the pure case at moderate temperatures we find that
the eigenvalue $\lambda$ scale as $1/M^2$, as is expected, while at the
lowest temperatures, the error due to the finite $m$ is larger making it 
difficult to see the expected $1/M^2$ scaling. For the impure
case, the convergence of $\lambda_{\mbox{\scriptsize imp}}$ with $M$ is more 
complicated, but the overall scaling is however still roughly $1/M^2$. 

We have also tested the convergence with the number of basis states
$m$. For both the pure and the impure cases we 
find a rapid convergence with increasing $m$. We have used a maximum of
$m=65$ for the calculations on the weakened link 
impurity and $m=38$ in the external spin case. We found however no
noticeable difference  between $m=38$ and $m=65$ down to $T=0.02$ in a test
run for the external spin. 

The truncation error, i.e.\ $1-\sum_{i=1}^m w_i$, where $w_i$ are the
largest eigenvalues of the density matrix, is less than about
$10^{-5}$ for $m=65$ at the lowest temperatures. Note that the truncation
error is determined during the pure sweep in which also the projection
operators and target states are determined. It could thus be used as an
estimate of an upper limit of the error of the pure properties, but it is
difficult to say how good estimate it yields for the impurity properties. 

To test our results we have also done Quantum Monte Carlo (QMC) simulations
for a few temperatures and couplings. The
DMRG data was well within the error bars of the QMC results. 

Local properties converge much faster with $m$ than the impurity
susceptibility. This fact
might be explained by the difficulty to numerically take a second
derivative, since we have to subtract two large numbers to find $\delta F$
in Eq.~(\ref{chiimp2}). This is however not the case for local 
properties, since we know that the local magnetization is zero in the
absence of a magnetic field. Another
source of inaccuracy in $\lambda_{\mbox{\scriptsize imp}}$ comes from the
error of the target 
states. Let us assume that the target state is determined up to some error
$\epsilon$; $|\psi \rangle_{\mbox{exact}} = | \psi
\rangle_{\mbox{\scriptsize DMRG}} + | \epsilon \rangle$. Since the target
state is,  to numerical accuracy, an eigenstate of ${T}_M$, the
eigenvalue will be  determined to order $\epsilon^2$. Expectation values of
other operators, for example the impurity transfer matrix, will however in
general only be accurate to order $\epsilon$. The local properties don't
seem to suffer too much from this effect,  the reason might be that
there is some cancellation of errors in the quotient $\langle \psi^L |
{T}_{\mbox{\scriptsize imp}}^{sz} | \psi^R \rangle / \lambda_{\rm imp}$.

While the accuracy of the second
derivative is good enough for the pure susceptibility down to about $T =
0.01$, we cannot trust the impurity susceptibility below temperatures roughly
an order of magnitude larger. Local impurity properties on the other hand
seem to be well represented down to about $T=0.02$. 

In summary we have shown that the transfer matrix DMRG is a useful method
for calculating finite temperature {\it impurity} properties of a spin chain
in 
the thermodynamic limit. We have considered two impurity
models: One weakened link and one external spin, but the method can be
applied 
to other impurity configurations and electron systems. We find that the
local response of 
the spin next to a weakened link always crosses over to a constant below
some $T_K$, i.e.\ to the behavior of the open chain fixed point. 
According to our calculations, the impurity susceptibility shows an exotic 
cross-over behavior with non-commuting limits. For the
external spin impurity, we have found that the data for the 
local response shows the expected cross-over to open chain behavior as the
temperature is 
lowered. The response has the scaling form in Eq.~(\ref{chifscaling}) and we
can explicitly show the data collapse and determine $T_K$ (Fig.~\ref{chif}).

\section{Acknowledgment}
We would like to thank H.~Johannesson, A.~Kl\"umper, T.~Nishino, I.~Peschel, 
S.~\"Ostlund, N.\ Shibata and X.\ Wang for 
valuable contributions. This research was supported in part by the 
Swedish Natural Science Research Council (NFR).

\end{document}